# Exploring Spin-Transfer-Torque Devices for Logic Applications

Zoha Pajouhi, Swagath Venkataramani, Karthik Yogendra, Anand Raghunathan and Kaushik Roy
Department of Electrical and Computer Engineering, Purdue University, West Lafayette, IN, USA
{zpajouhi,venkata0,kyogendr,raghunathan,kaushik}@purdue.edu

*Abstract*-As CMOS nears the end of the projected scaling roadmap, significant effort has been devoted to the search for new materials and devices that can realize memory and logic. Spintronics, which uses the spin of electrons to represent and manipulate information, is one of the promising directions for the post-CMOS era. While the potential of spintronic memories is relatively well known, realizing logic remains an open and critical challenge. All Spin Logic (ASL) is a recently proposed logic style that realizes Boolean logic using spin-transfer-torque (STT) devices based on the principle of non-local spin torque. ASL has advantages such as density, non-volatility, and low operating voltage. However, it also suffers from drawbacks such as low speed and static power dissipation. Recent work has shown that, in the context of simple arithmetic circuits (adders, multipliers), the efficiency of ASL can be greatly improved using techniques that utilize its unique characteristics. An evaluation of ASL across a broad range of circuits, considering the known optimization techniques, is an important next step in determining its viability.

In this work, we propose a systematic methodology for the synthesis of ASL circuits. Our methodology performs various optimizations that benefit ASL, such as intra-cycle power gating, stacking of ASL nanomagnets, and fine-grained logic pipelining. We utilize the proposed methodology to evaluate the suitability of ASL implementations for a wide range of benchmarks, *viz.* random combinational and sequential logic, digital signal processing circuits, and the Leon SPARC3 general-purpose processor. Based on our evaluation, we identify (i) the large current requirement of nanomagnets at fast switching speeds, (ii) the static power dissipation in the all-metallic devices, and (iii) the short spin flip length in interconnects as key bottlenecks that limit the competitiveness of ASL. We further evaluate the impact of various potential improvements in device parameters on the efficiency of ASL.

## I. Introduction

As CMOS devices scale down to the deep nanometer regime and approach their fundamental physical limits, the traditional benefits in power and performance associated with technology scaling have subsided due to increased short-channel effects and leakage power consumption [1]. This has motivated researchers to explore newer devices that can potentially replace CMOS as the next Boolean "switch" [2-4]. Specifically, recent advances [5-11] and experiments [12-14] on spin-transfer-torque (STT) devices have identified the possibility of using "spin" (rather than "charge") as the state variable for computation. STT devices manipulate the spin orientation of a nanomagnet using a spin-polarized current to switch between Boolean logic states. These devices possess several desirable characteristics: (i) they are *non-volatile*, since the spin orientation is retained in the nanomagnet even when the power supply is turned off, (ii) they offer high *density*, and (iii) they can be operated at very low voltages in the order of 10 mV. Due to these characteristics, STT devices have been extensively explored and demonstrated to be efficient in the design of both on-chip [15-17] and off-chip [18-20] memories.

The intrinsic benefits of STT devices can also be potentially leveraged in the context of logic design [21-24]. Towards this end, All Spin Logic (ASL) [25,26] is a recently proposed approach to implement Boolean logic functions using spin-based devices. ASL circuits are comprised of a network of nanomagnets, which represent the internal logic signals of the circuit, interconnected through non-magnetic metallic channels. We illustrate the principle and operation of ASL through the example shown in Figure 1 (a). The circuit consists of 2 nanomagnets - an injecting or input nanomagnet (*M1*) and receiving or output nanomagnet (*M2*)- connected through a spin channel. This 1-input 1-output circuit can operate as either a buffer or an inverter as explained below. Let us assume that the nanomagnets are initially polarized in opposite directions - *M1* to the right and *M2* to the left. Now, when a current is passed from top to bottom through *M1*, the nanomagnet acts as a *polarizer* and polarizes the spin of the electrons parallel to its orientation (right-spin in this case). This spin-polarized current travels through the channel and exerts a spin torque on *M2* based on the principle of non-local spin torque (NLS) [11, 12]. If the exerted spin torque is strong enough, the orientation of *M2* is reversed to align with *M1*. Thus, by passing a current through *M1*, its spin orientation is transferred to *M2*, thereby realizing the functionality of an inverter. Interestingly, the circuit can also operate as a buffer if the direction of current through *M1* is reversed *i.e.,* the current is passed from bottom to top as shown in Figure 1 (b). In this case, *M1* polarizes the current in a direction opposite to its orientation resulting in the channel current, and hence *M2,* to orient opposite to *M1*. Thus the circuit can either "copy" or "invert" the orientation of the input (*M1)* to the output (*M2)* based on the direction of current through *M1*.

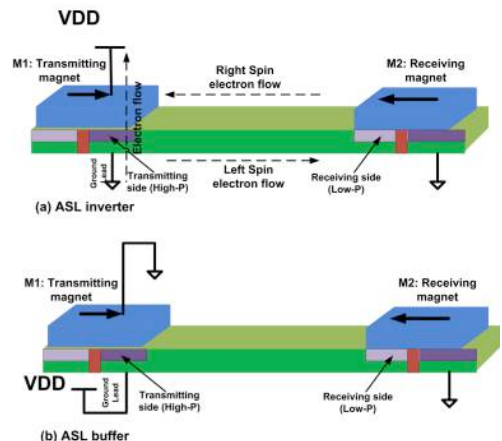

Fig. 1: (a) ASL inverter and (b) ASL buffer.

Recent efforts [7, 10] have adopted a similar approach to design complex ASL gates such as AND, OR, XOR *etc*. Also, the design of a compact full-adder using ASL has been proposed [11], which has in turn been used to design and evaluate an ASL implementation of the Discrete Cosine Transform (DCT). While the above efforts have made a promising start, the suitability of ASL or the realization of larger and a broader range of logic circuits remains unexplored. To enable such an exploration, we propose a systematic methodology to synthesize arbitrary logic circuits using ASL. Our methodology incorporates various optimizations that exploit the unique properties of ASL. First, we exploit the non-volatility of nanomagnets to realize storage elements with minimal cost. Second, power is consumed in the all-metallic ASL gates regardless of whether or when any useful switching occurs. Thus, power gating or "clocking" ASL gates in a fine-grained manner is critical to avoid significant energy penalties due to leakage. However, power gating incurs area and energy overheads in the form of gating transistors, thus requiring proper analysis and optimization. Towards this end, our methodology automatically identifies ASL gates with similar time-periods of evaluation and clusters them into gating domains. ASL gates within each gating domain share a gating transistor, thereby amortizing the overheads of these transistors while retaining a significant fraction of the energy benefits.

Another avenue for energy optimization in ASL circuits is "magnet stacking" [31], which refers to connecting the terminals of multiple nanomagnets in series. Through manual design of arithmetic circuits, stacking has been shown to significantly improve the energy efficiency of ASL [31]. Finally, a key consideration in designing ASL circuits is that, since ASL gates communication using spin-current, the physical length of interconnects cannot exceed the so-called spin diffusion length (*λsf*) or spin flip length of the communication channel. Thus, buffers should be inserted in larger interconnects to ensure correct functionality. Finally, taking advantage of the built-in non-volatility, ASL circuits can be operated in a pipelined fashion, improving their throughput.In summary, the key contributions of this work are:

- We propose the first systematic methodology for implementing arbitrary logic circuits using ASL.
- The proposed methodology incorporates optimizations such as intra-cycle power gating, fine-grained logic pipelining, and nanomagnet stacking, which exploit the unique properties of STT devices.
- We utilize the methodology to evaluate ASL for implementing combinational and sequential logic benchmarks, DSP circuits, and the Leon SPARC3 general-purpose processor.
- Based on our evaluation, we draw key conclusions about the viability and competitiveness of ASL for different performance scenarios. We also evaluate the impact of potential improvements in key material and device parameters on the efficiency of ASL.

The rest of the paper is organized as follows. Section II provides background on ASL circuits and their characteristics. Section III outlines the proposed ASL synthesis methodology and the techniques that it employs. Section IV presents the evaluations performed using the synthesis methodology and discusses the results obtained for the different benchmarks. Finally, Section V concludes the paper.

II. ALL SPIN LOGIC: PRELIMINARIES

All Spin Logic operates on the basic principle of storing state information in the spin of electrons (magnets) and manipulating the state using spin-polarized currents. An ASL buffer (inverter) consists of two nanomagnets that are connected through a non-magnetic and metallic channel as shown in Figure 1. In order to understand the operation of the buffer (inverter), we will distinguish between "charge current" and "spin current". The input nanomagnet, *M1*, (connected to a supply voltage) injects spin-polarized electrons into the channel (which is made of a non-magnetic material such as Cu). Even though the charge current flows from supply to ground, spin-polarized current can flow through the channel due to the spin potential difference across it. The spin current exerts a torque to flip the output nanomagnet, *M2*. Note that, in Fig. 1 (a) the left-spin electrons flow to the right and right-spin to the left in the channel, making the total charge current to be zero. Since the channel is non-magnetic, the spin orientations of electrons would get randomized beyond the spin flip length of Cu, the spin flip or *λsf* is about 500nm [27]. In other words, if the input and output nanomagnets are separated by more than 500nm, either there is a need for repeaters, or it will not be possible to switch the output nanomagnet, *M2*, using spin-polarized current injected from nanomagnet *M1*. The high and low polarized layers placed below the nanomagnets guarantee the directionality of the logic gate.

To simulate an ASL gate, each of the gate's elements such as the nanomagnets and the channel are modeled as four component conductance elements: one charge conductance and three spin conductances [7,10]. Then, the current is derived and inserted into a magnetization solver. The magnetization dynamics are determined by self-consistently solving the current equation using the spin conductances and the Landau-Lifshitz-Gilbert (LLG) equation [28]. A more detailed description of the simulation methodology can be found in Section III A.

The basic ASL inverter can be extended into a majority gate as shown in Figure 2, where *A*, *B* and *C* are the input nanomagnets and *D* is the output nanomagnet. Majority gates can be used to realize other logic functions e. g. if one of the nanomagnets in Figure 2 is fixed to logic `1', the gate computes the NOR of the remaining inputs. It is of particular interest that each nanomagnet can be considered as a storage element. Hence, if properly designed, there is no need for separate flip-flops or latches in ASL circuits.

Since ASL devices are all metallic, they are capable of functioning at very low voltages (~10mV) [28,31]. However, they have a poor performance in comparison with CMOS. Improved performance can be achieved by injecting larger current through the nanomagnets [28,29]. However, increasing the current has a detrimental effect on the power consumption and can cause reliability concerns.

Unlike CMOS circuits in which power consumption is higher during switching time and lower at other times, ASL circuits consume similar power regardless of whether they are switching (due to the metallic direct path from supply to ground). Therefore, there is a need for "power gating" the devices with an added transistor to reduce power dissipation when the gate is not evaluating. Figure 3 depicts two cascaded gates with power gating. The transistors are used for powering the ASL nanomagnets. For example, *M1* is enabled through *T1* and if *T1* is turned on, inverter 1 will be evaluated and *M2* flips according to its input. At the next step, *M1* can be switched off and *M2* can be turned on; thus, inverter 1 is turned off and inverter 2 is turned on. At this step, the new magnetization of *M2* gets propagated to *M3*.

The ASL gate can be viewed as a resistor whose resistance is based on the size of the nanomagnet (typically a few Ohms) in series with the MOSFET. The voltage across the source to drain of the transistor can be of the order of few tens of millivolts, mainly determined by the energy restrictions (note that the gating transistor operates in the triode region). However, since the resistance of the triode region transistor is expected to be much larger (in the range of KOhms) than the resistance of the nanomagnet (few Ohms), most of the power dissipation will be in the gating transistor. In order to mitigate this drawback, one can stack several nanomagnets and share one gating transistor for all of them (note that this can only be done when the nanomagnets are evaluated simultaneously). This enhances power efficiency because the power consumption of the MOSFET is amortized across the nanomagnets sharing that transistor. Figure 4 shows a sample circuit with stacking. It can be observed that since *A*, *B* and *C* are inputs to the same gate and are active simultaneously, T1 can be used to power gate all nanomagnets corresponding to *A*, *B* and *C*.

Thus, in order to evaluate the potential of ASL, there is a need to utilize suitable design techniques including power gating and stacking of ASL gates. In addition, as it will be described in the next section, the performance of ASL circuits can be significantly improved through fine-grained pipelining by exploiting the built-in sequential elements in ASL gates. We next describe a synthesis methodology for ASL that incorporates these optimizations.

## III. ASL SYNTHESIS METHODOLOGY

In this section, we explain the proposed synthesis methodology for ASL. The methodology takes an RTL description of the circuit as its input and produces an ASL implementation (ASL gate netlist with the required gating transistors) that is optimized for energy. Figure 5 shows the various steps involved in the synthesis process. First, a standard cell technology library is developed for ASL, in which different logic gates are optimized and characterized using a physics-based simulation framework. Next, a commercial logic synthesis tool is utilized to perform technology-independent optimization and a basic mapping of the circuit to the ASL technology library. Several design optimizations are then applied to enhance the energy-efficiency of the synthesized netlist. These include: (i) *Intra-cycle power gating*, in which the nanomagnets in the ASL netlist are clustered into multiple gating domains, each of which are dynamically power gated during circuit operation, (ii) *Nanomagnet stacking,* in which the nanomagnets within each gating domain are further sub-divided and their terminals are "stacked" together in series, and (iii) *Fine grained logic pipelining*, in which the logic paths within each gating domain are balanced through the insertion of buffers and are operated in a pipelined fashion. We note that the above optimizations incur overheads in terms of both area and energy and hence require careful analysis of the trade-offs involved. In addition, to ensure correct operation, our synthesis methodology also performs fanout optimization and approximate placement followed by interconnect buffer insertion. The following subsections provide a detailed description of the above steps.

### A. ASL Technology Library Generation

The ASL technology library comprising of a range of logic gates with varying energy-delay characteristics is developed using a rigorous physics-based simulation framework. A physics-based simulation framework for ASL analysis was proposed in [10]. To simulate ASL devices, the *spin transport* through the devices and the *magnetization dynamics* have to be solved self-consistently. The transport simulation is based on a modified Valet-Fert model [30] and the magnetization

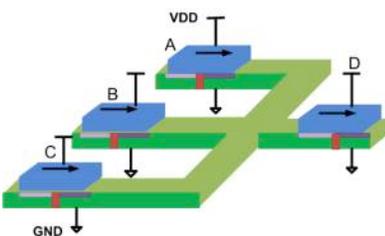
Fig. 2: ASL majority gate circuit.

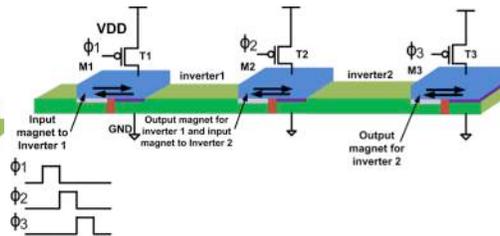
Fig. 3: ASL power gating circuit.

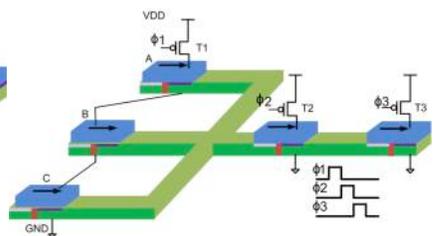
Fig. 4: Sharing/stacking of nanomagnets.

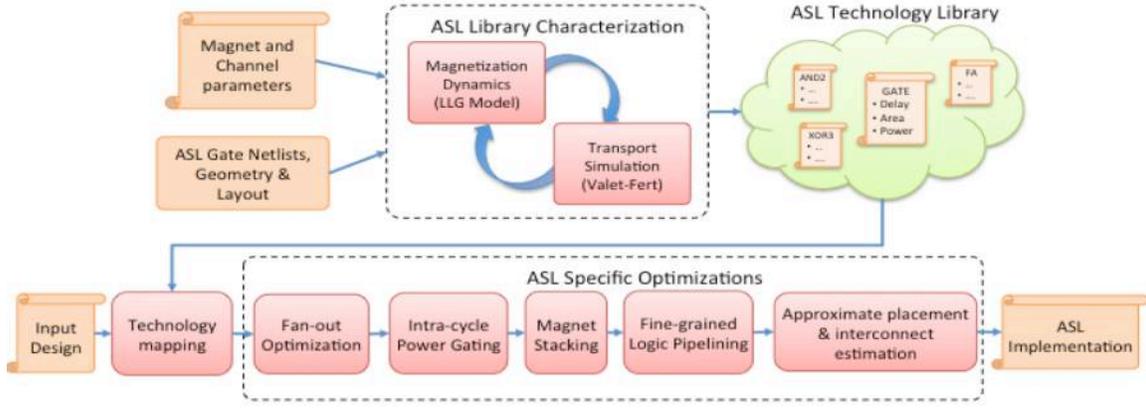

Fig. 5: ASL synthesis methodology diagram.

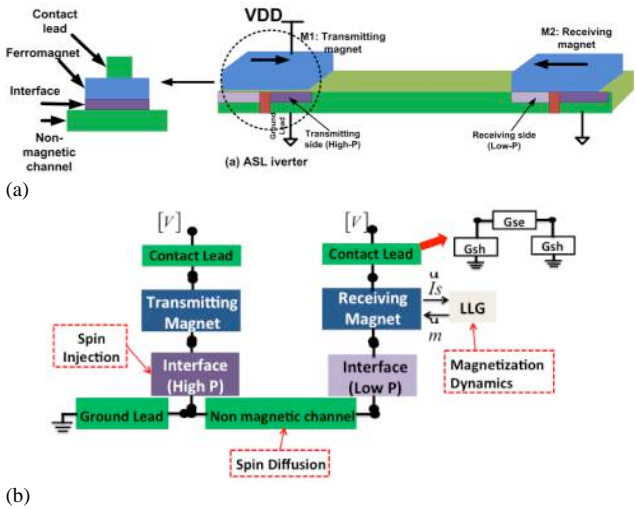

Fig. 6: (a) Schematic of buffer/inverter in ASL (b) The different building blocks/components used to model ASL gate [31].

dynamics are computed using the Landau-Lifshitz-Gilbert (LLG) equation with spin-torque [29]. The transport simulation model receives the magnetization vector ($\hat{m} = M_x, M_y, M_z$) of the nanomagnets and calculates the current in the channel as a 4-component vector ($I_{4x1} = [I_c, I_s^x, I_s^y, I_s^z]$). The channel current is then fed to the magnetization dynamics model, which provides the magnetization vector as the output. Thus, the magnetization dynamics is a function of spin currents, which in turn, are estimated based on the spin transport simulations. On the other hand, the transport calculations depend upon the direction of spin momentum in the nanomagnet and the spin channel. Hence the models are solved self-consistently to accurately estimate the switching time and energy of ASL gates [31].

From a modeling perspective, the ASL device can be divided into 4 regions as shown in Figure 6a: (a) non-magnetic channel to transport the charge current from the contact lead to the nanomagnet, (b) ferromagnet that acts as a source of spin polarized electrons, (c) an interface region between ferromagnet and non-magnetic channel that enhances the injection of spins from the nanomagnet into the channel and (d) non-magnetic channel to transport spin current from the input nanomagnet to the output. The connections between these four regions are shown in Figure 6 (b). Each of these 4 regions is modeled using a lumped π-conductance model with a series element (Gse) and two shunt elements (Gsh). Since spin current is modeled as a vector, all the nodal voltages and branch currents of the circuit are represented using four components ([Vc,Vz,Vx,Vy] and [Ic,Iz,Ix,Iy]) – one element for charge and 3 directional elements for spin. All the conductances (G) are represented by 4x4 matrices which connect the branch currents with nodal voltages as per Ohm's law:

$$[I_c \quad I_z \quad I_x \quad I_y]^T = G * [V_c \quad V_z \quad V_x \quad V_y]^T \quad (1)$$

The details of different conductance matrices are given below:

*(A) Non-magnetic channel*

The spin current from transmitting nanomagnet diffuses to the receiving nanomagnet through non-magnetic (NM) channel. The lumped series (Gse_ch) and shunt (Gsh_ch) conductance elements of non-magnetic channel are as represented below.

$$Gse\_ch = \frac{A}{\rho L} * eye\left(1, \left(\frac{L}{\lambda_{sf}}\right) \operatorname{csch}\left(\frac{L}{\lambda_{sf}}\right), \left(\frac{L}{\lambda_{sf}}\right) \operatorname{csch}\left(\frac{L}{\lambda_{sf}}\right), \left(\frac{L}{\lambda_{sf}}\right) \operatorname{csch}\left(\frac{L}{\lambda_{sf}}\right)\right) \quad (2)$$

$$Gsh\_ch = \frac{A}{\rho L} * eye\left(0, \left(\frac{L}{\lambda_{sf}}\right) \tanh\left(\frac{L}{2\lambda_{sf}}\right), \left(\frac{L}{\lambda_{sf}}\right) \tanh\left(\frac{L}{2\lambda_{sf}}\right), \left(\frac{L}{\lambda_{sf}}\right) \tanh\left(\frac{L}{2\lambda_{sf}}\right)\right) \quad (3)$$

Here, *eye(a1,a2,a3,a4)* is an eye matrix with all non-diagonal coefficients equal to zero and the diagonal coefficients equal to *a1* to *a4*. *L* is the length of the channel, *A* is the area of the cross section, *ρ* is the resistivity and *λsf* is the spin diffusion length. The first element of the series component is the charge conductance arising from ohm's law and the other elements are spin diffusion terms. Note that the first element of the shunt component is zero indicating that the shunt component only acts as a spin sink.

*(B) Ferromagnetic (bulk) region*

The four component lumped conductance matrices for a nanomagnet aligned along the 'z' direction is given by:

$$Gse\_FM = \begin{bmatrix} G11 & Z \\ Z & Z \end{bmatrix}, G11 = \frac{A}{\rho L}\begin{bmatrix} 1 & P \\ P & P^2 + \alpha \operatorname{csch}\left(\frac{L}{\lambda_{sf}}\right) \end{bmatrix},$$
$$Z = \begin{bmatrix} 0 & 0 \\ 0 & 0 \end{bmatrix} \quad (4)$$

$$Gsh\_FM = \begin{bmatrix} G11 & Z \\ Z & Z \end{bmatrix}, G11 = \frac{A}{\rho L}\begin{bmatrix} 0 & 0 \\ 0 & \alpha \tanh\left(\frac{L}{2\lambda_{sf}}\right) \end{bmatrix},$$
$$Z = \begin{bmatrix} 0 & 0 \\ 0 & 0 \end{bmatrix} \quad (5)$$

Where $\alpha = 1 - P^2 L\lambda sf$ and P is the spin polarization factor.

*(C) Channel-Magnet interface*

The series and shunt components of the interface region consider the mode mismatch between ferromagnet and the non-magnetic channel.

$$Gse\_int = \begin{bmatrix} G11 & Z \\ Z & Z \end{bmatrix}, G11 = \frac{q^2}{h}M\begin{bmatrix} 1 & P \\ P & 1 \end{bmatrix}, Z = \begin{bmatrix} 0 & 0 \\ 0 & 0 \end{bmatrix} \quad (6)$$

$$Gsh\_int = \begin{bmatrix} Z & Z \\ Z & G22 \end{bmatrix}, Z = \begin{bmatrix} 0 & 0 \\ 0 & 0 \end{bmatrix}, G22 = \frac{q^2}{h}M\begin{bmatrix} a & b \\ -b & a \end{bmatrix} \quad (7)$$

For ohmic contacts a~1 and b~0. For circuit simulation, the ASL device is divided into different building blocks/components connected to each other with each component represented by a lumped π conductance model. The magnetization dynamics of the output nanomagnet is obtained by solving the Landau-Lifshitz-Gilbert (LLG) equation:

$$(1+\alpha^2)\frac{d\hat{m}}{dt} = -|\gamma|(\hat{m}\times\vec{H}) - \alpha|\gamma|(\hat{m}\times\hat{m}\times\vec{H}) + \vec{\tau} + \alpha(\hat{m}\times\vec{\tau}) \quad (8)$$

Where $\vec{\tau} = \frac{\hat{m}\times\vec{Is}\times\hat{m}}{qNs}$ is the spin torque. Here α is the Gilbert damping constant, $\hat{m}$ is the magnetization of the receiving or output nanomagnet, γ is the gyromagnetic ratio, q is the charge of an electron, $Ns$ is the total number of spins in the nanomagnet given by the relation $Ns = \frac{M_s\Omega}{\mu_B}$ (Ms- Saturation magnetization, Ω-Volume of the nanomagnet and $\mu_B$ is the Bohr magneton). $\vec{H}$ is the effective field which represents the sum of internal and external fields on the nanomagnet. Figure 7 shows the magnetization dynamics of 'z' component for Iin=500μA and Iin=5mA. As expected the nanomagnet can be switched faster with higher input current; however, it leads to higher power dissipation.

Utilizing the simulation framework described above, ASL gates of different functionality can be designed and characterized. The ASL library developed in this work contains 30 logic gates with a maximum fan-in size of 7. There are several design parameters that affect the delay and energy characteristics of ASL gate. The first parameter is the ground resistance. In ASL gates, only a portion of the input spin current flows through the spin channel to the output nanomagnet, while the rest of it is lost to the ground. The fraction of the spin current that flows to the ground can be manipulated by varying the ground lead resistance ($R_g$) of the gate. Thus, the ground resistance has a direct impact on the delay characteristics of the ASL gate. As shown in Figure 8, as the ground resistance is increased, a larger fraction of input spin current reaches the output nanomagnet and hence the delay of the gate decreases. Note that the charge component of the input current completely flows to the ground as the output nanomagnet in the ASL gate is floating. As described in Section II, fine-grained power gating is key to the energy efficiency of ASL. In order to power gate the device, one has to connect it to a transistor. The gating transistor should be designed to supply the required current ($I_{ON}$) during the evaluation period. For proper operation, the gating transistors require a substantially higher power supply in comparison with the power supply required for ASL gate operation. This large power supply is due to the $R_{ds}$ of the transistor. However, upsizing the transistor increases its dynamic power. Therefore, there is a trade-off between the dynamic and static power overhead of the gating transistor. Figure 9 shows the static and dynamic power consumption of an inverter versus the normalized area of the transistors for a constant $I_{ON}$. It can be observed that the static power decreases drastically and the dynamic power increases with an increase in the area of the transistor.

### B. Technology Mapping and Fan-out Optimizations

Once the ASL library is generated, utilize a commercial synthesis tool (e.g. Synopsys Design Compiler [32]) to map the input RTL to the ASL library.

While this yields a functional map to the standard cells in the ASL library, additional optimizations are required to ensure correct operation due to the following circuit-level considerations: (i) The fan-out of ASL gates cannot exceed 1. This is because the spin current in the channel is sufficient to influence only one receiving or output nanomagnet. (ii) The length of interconnects or spin channels between successive gates cannot exceed the *λsf* of the material.

We propose two different methods *viz.* performance optimized fan-out and power optimized fan-out, to achieve proper operation for gates with larger fan-outs. In the case of performance-optimized fan-out, as shown in Figure 10 (a), the number of receiving nanomagnets is increased to the number

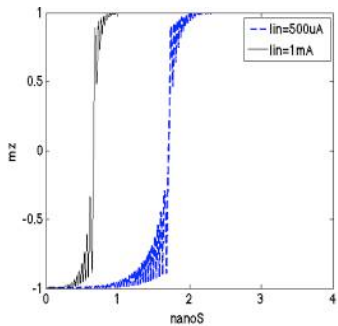
Fig. 7: Magnetization dynamics of ASL for Iin=500μA and 1mA

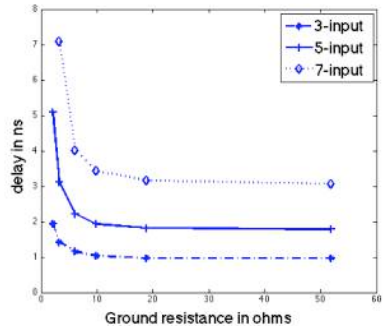
Fig. 8: Switching delay *vs.* ground resistance for different ASL gates.

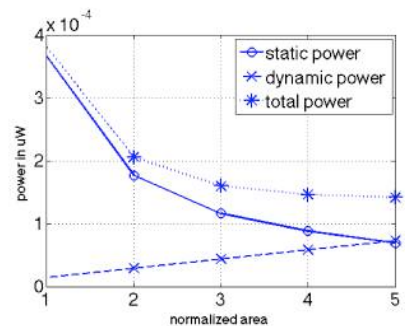
Fig. 9: Power *vs.* relative area for an inverter.

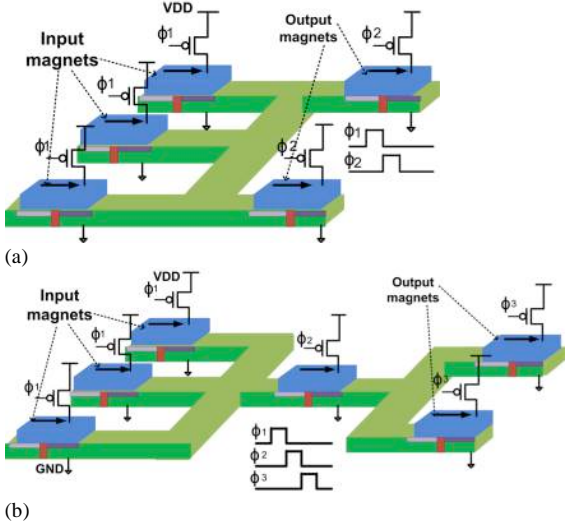

(a)

(b)

Fig. 10: (a) Performance optimized fan-out. (b) Power optimized fan-out.

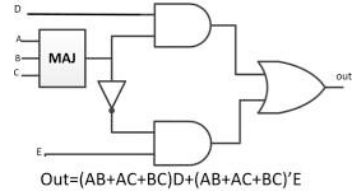

Fig. 11: Sample benchmark---gate level implementation.

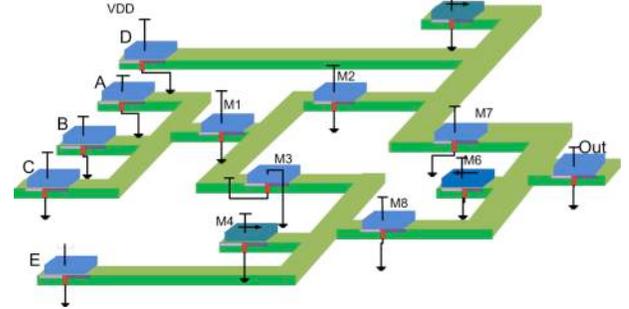

Fig. 12: Gate netlist for the sample benchmark.

of fan-outs of the gate (2 in this example). However, the gating transistors of all the input nanomagnets are sized up such that the current through the spin channel is sufficient enough to flip all the receiving nanomagnets. Clearly, this incurs area and power overheads, but does not result in any performance degradation.

On the other hand, as shown in Figure 10 (b), the power-optimized fan-out evaluates the gate in two stages. In the first stage, the logic is evaluated by supplying current through the input nanomagnets and the output is stored in an intermediate receiving nanomagnet. In the second stage, a larger current, sufficient to flip all the output nanomagnets, is passed through the intermediate nanomagnet and its spin orientation is transferred to the outputs. Since power-optimized fan-out introduces an extra stage of logic, it potentially degrades performance if the gate lies in the critical path. However, it alleviates power overheads, as a larger current is only supplied to the intermediate nanomagnet, as opposed to all the input nanomagnets in the previous case. In our methodology, we perform power-optimized fan-out if there exists timing slack at the output of the gate; else we resort to performance-optimized fan-out.

The the above steps provide a fully functional ASL implementation of the given circuit. Next, we perform several optimizations to improve the energy consumption of the circuit, which are detailed in the following subsections. We illustrate different optimizations using an example circuit shown in Figure 11. The corresponding ASL gate netlist obtained after technology mapping including power-optimized fan-out is shown in Figure 12. Note that M4-M6 in Figure 12 are fixed nanomagnets used to realize different logic functions from majority gates.

### C. Generation of Power Gating Domains

One of the key differences between CMOS and ASL is that the energy consumption of the circuit is not data-dependent, *i.e.*, energy is consumed when current is supplied to the input nanomagnets of a gate, regardless of whether the output nanomagnet changes its spin orientation. Hence, it is critical to turn-off the current supply (by switching-off the gating transistor) to the input nanomagnets as soon as the evaluation of the gate is complete. This is possible because the nanomagnets are non-volatile and they retain their state even when the gating transistor is turned off.

Power gating each gate individually will incur significant overheads in terms of the logic required to generate the power gating signals. Instead, we cluster the gates in the circuit into several "gating domains" and gates within each domain are evaluated and power-gated together. Now, each gating domain should be powered ON for the duration equal to the union of the evaluation periods of constituent gates. Clearly, this duration is greater compared to when each gate was power-gated individually. Therefore, while clustering reduces the energy overheads associated with generating the power-gating signals, it negatively impacts the evaluation energy of the gating domains, as the gates are potentially powered ON for a longer period of time. An extreme example is when all gates in the circuit are clustered to a single gating domain, in which case they are powered ON for the entire gating period resulting in significant energy penalties. Thus, the circuit should be carefully clustered such that the number of gating domains is minimized while ensuring the evaluation period of the gates is minimum.

Algorithm 1 shows the pseudo-code used to generate the gating domains. The algorithm is based on searching to obtain a local minimum for the energy consumption [33]. The algorithm takes an input gate netlist and the pulse width of the multiphase clock gating as its inputs and produces a clustered netlist with the logic necessary to power-gate the gating transistors. First, the circuit is levelized and the gates are sorted in topological order. Next, gates within the same logic level are clustered together to form separate gating domains. The intuition behind this clustering is that gates in the same level of logic have similar arrival times. This initial cluster configuration is further improved such that the overall energy is minimized. We specifically consider the critical gates in each cluster, *i.e.*, the gates at the boundaries of the cluster that determine the time duration for which the cluster should be powered ON. We perturb the cluster configuration by moving the critical gates between clusters. For example, consider the case where gate *G* originally in cluster *i* is moved to cluster *j*. This results in the ON duration of cluster *i* reducing by $\Delta T_D$ and potentially increases the ON duration of cluster by *j* by $\Delta T_I$. Also, let $N_i$ and $N_j$ be the number of gates in the clusters *i, j* respectively. Then, we estimate relative energy savings when the gate is transferred to be the difference between the product of the number of gates in the cluster and the change in its time duration ($\Delta T_I . N_j - \Delta T_D . N_i$). If this value is beyond a threshold, then the gate is transferred from cluster *i* to *j*. This process is iterated until no gate movement changes the clusters and the final cluster configuration is thus obtained.

Given the cluster configuration, the power gating signals for each cluster are generated using a multi-phase clock. The clusters are sorted in topological order and based on the time duration, for which they are required to be powered ON, appropriate number of gating phases are assigned. Thus intra-cycle power gating is achieved. Figure 13 shows the ASL implementation of the sample benchmark with power gating transistors added. In this case, the sample benchmark has 5 gating domains for the different levels of logic in the circuit. As seen, A,B,C are assigned to the gating domain 1, M1 is assigned to gating domain 2, D,E and M2-M5 are assigned to domain 3, M6-M8 are assigned to domain 4 and finally the output nanomagnet is assigned to domain 5.

### D. *Magnet Stacking*

Another key avenue for energy optimization in ASL circuits is to share the gating transistors amongst gates within a gating domain, since they are evaluated and power-gated together during circuit operation. At the circuit level, each nanomagnet can be modeled as a resistor, with resistances as low as 2Ω to 30Ω. The key idea is to connect the nanomagnets (both within and across gates) in series with the gating transistor. This arrangement is referred to as "stacking". Stacking nanomagnets results in significant power improvements as

---

**Algorithm 1** Generate Gating Domains for Intra-cycle Power Gating

*Input*: Circuit (CKT), Multi-phase Gating
*Output*: Gating domains {1,2,3, … K}
         Gates in each gating domain {$C_1$, $C_2$, … $C_K$}
1. Levelize CKT
2. Assign gates in Level *i* → clock domain $C_i$
3. {$N_1$, $N_2$, … $N_K$} = Number of gates in each domain
4. $\Delta M = \infty$
5. **While** ($\Delta M > 0$) {
6.     $\Delta M = 0$
7.     **For** i = 1 to K   % each clock domain
8.         $G_S$ = {Gates that determine *start* time of domain i}
9.         $G_E$ = {Gates that determine *end* time of domain i}
10.        $\Delta T_D$ = Time decrease of i when $G_S$ moved to i-1
11.        $\Delta T_I$ = Time increase of i-1 when $G_S$ moved from i
12.        **If** ($\Delta T_I . N_{i-1} - \Delta T_D . N_i$) {
13.           Move $G_S$ → domain i-1
14.           $\Delta M$ += $G_S$
15.           $N_i$ -= $G_S$ ; $N_{i-1}$ += $G_S$
16.        } % End if
17.        Repeat lines 10-15 by moving $G_E$ to domain i+1
18.        **If** ($N_i == 0$) % if all gates moved to adjacent domains
19.           Delete domain; K -= 1
20.     } % End for
21. } % End while
22. No of Clock phases = K
23. Assign clock domain *i* to clock phase i

---

follows. Let us assume the simple case of gating without any stacking involved. Under this assumption, the number of gating transistors would be equal to the number of nanomagnets. Let $V_A$ be the operating voltage of the ASL nanomagnets and $V_C$ denote the voltage drop in the transistor. Then, a first order approximation of the static power *(P)* consumed by the ASL circuit is given by Equation 9.

$$P = (V_C + V_A). N. I \qquad (9)$$

In the above equation, *N* represents the number of nanomagnets (or gating transistors) in the circuit and *I* denotes the current. In the ideal case of $V_C \approx 0$ there is no need for stacking because the overhead of the transistor is negligible. However, due to the ON resistance of the gating transistors, the overhead of the transistors is large. Therefore, $V_C \gg V_A$ and hence, the power consumption of the circuit is dominated by the gating transistor.

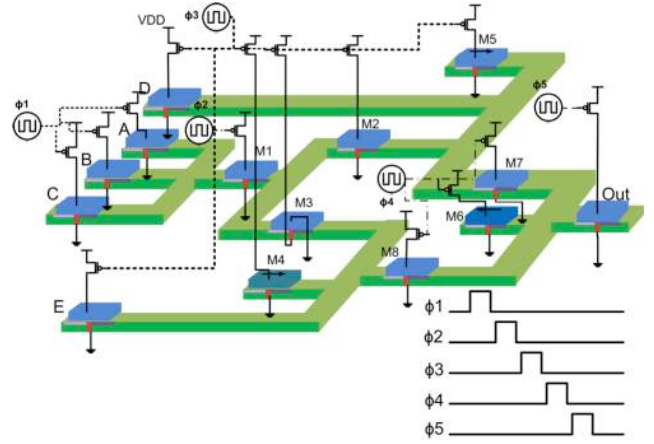

Fig. 13: Gating domains for intra-cycle power gating for a sample benchmark.

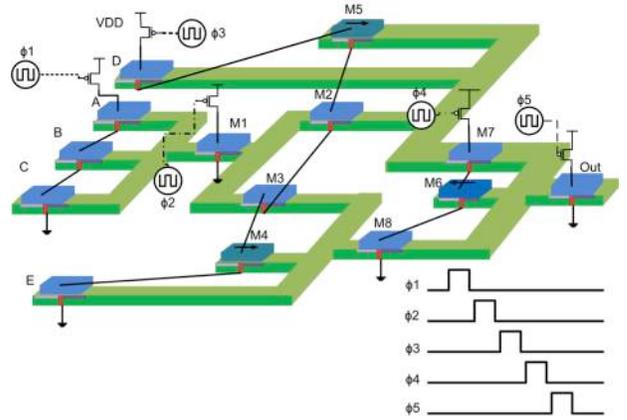

Fig. 14: Magnet stacking for the sample benchmark

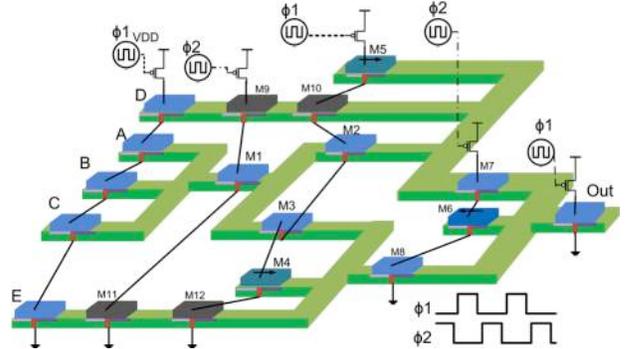

Fig. 15: Fine grained logic pipelining for the sample benchmark.

Now, let us assume that the network topology allows stacking of '$m$' nanomagnets. In this case, the power consumed ($P_{STACK}$) is given by Equation 10.

$$P_{STACK} = (V_C + m.V_A) . (N/m) . I \qquad (10)$$

Since the nanomagnets are connected in series, the supply voltage is increased to $V_C+m.V_A$. However, the relative increase in the supply voltage is insignificant for small values of '$m$' as $V_C \gg V_A$. On the other hand, the number of gating transistors decreases linearly with '$m$', resulting in an almost linear decrease in the power consumption of the circuit. Note that the current requirement does not increase when the nanomagnets are stacked as they are connected in series with each other. Stacking increases the complexity of routing, as the nanomagnets need to be connected in series with the gating transistor. Hence the degree of stacking ($m$) is primarily constrained by the physical limitations of routing. The nanomagnet stacking optimization for the sample benchmark of Figure 11 is illustrated in Figure 14. Note that the nanomagnets in each gating domain are connected in series with the source terminal of the corresponding power gating transistor.

*E. Fine-grained Logic Pipelining*

Logic pipelining is a common optimization used to improve throughput. In conventional CMOS circuits, logic pipelining incurs significant power overheads in the form of latches or flip-flops. In contrast, the nanomagnets in ASL may themselves be viewed as state elements and therefore the circuits can be pipelined in a fine-grained manner with minimal overhead. The gating domains identified in Section III C can be utilized for this purpose. The key idea is that once a gating domain is evaluated on a given input, instead of power gating the domain, it can alternatively be used to evaluate subsequent inputs in a pipelined manner. However, this requires several design considerations. First, successive pipeline stages cannot be operated together, as the output nanomagnets of the first stage are the input nanomagnets of the next. Hence we apply a two-phase pipelining strategy wherein only the odd stages are operated in one cycle, followed by the even stages in the next.

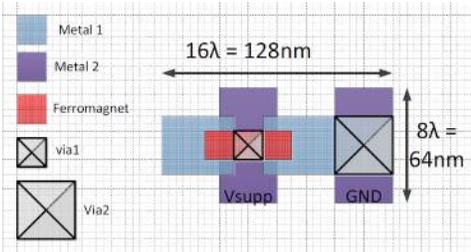

Fig. 16: Layout of an ASL nanomagnet.

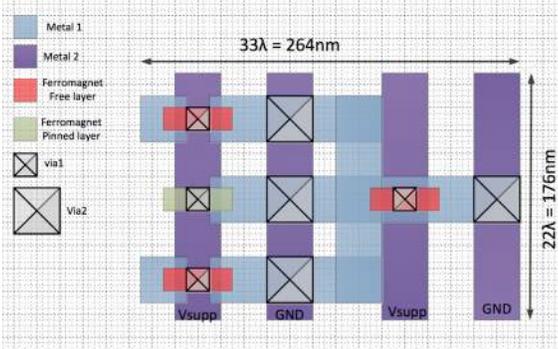

Fig. 17: Layout of an ASL NAND gate.

Another constraint for logic pipelining is that the paths in the circuit should be balanced, *i.e.*, if there exists a connection between gates that are not located on adjacent gating domains (or pipeline stages), then buffers need to be inserted in each intermediate gating domain to latch the state, as it will be overwritten in the course of pipelined operation. The buffer insertion can be expensive depending on the number of fan-ins and fan-outs for a given gate. Hence, we utilize a modified list scheduling [34-36] procedure to move gates across gating domains to minimize the buffer costs. Note that, while logic pipelining significantly improves throughput of the circuit, it may not be feasible when sequential dependencies exist across different pipeline stages. The proposed methodology aggressively pipelines the circuit while ensuring such dependencies are not violated. Figure 15 shows the ASL implementation of the benchmark in Figure 11 with fine-grained logic pipelining. Nanomagnets M9-M11 in the netlist are inserted to balance the paths in the design in order to facilitate pipelined operation.

*F. Approximate Placement and Interconnect Estimation to Determine Buffers/Repeaters*

All Spin Logic operation is based on spin current injection through an input nanomagnet, diffusion of the spin current through the metallic channel, and updating the state of the output nanomagnet. However, $\lambda sf$ of the channel material (Cu in our case) limits the flow of spin current in the channel. Therefore, if the channel length is longer than the $\lambda sf$, there is a need to insert one or more buffers or repeaters (nanomagnets). Estimating the number of buffers requires a detailed place-and-route of the design, considering various physical design rules. In this work, we adopt an approximate placement methodology and estimate interconnect lengths based on this placement as described below.

Each ASL gate consists of nanomagnets communicating through Cu channels and the corresponding gating transistor(s). Note that there are two types of interconnects in ASL: (a) charge based interconnects for the gating transistors and (b) spin based interconnects (channels) for ASL logic operation, which are limited by $\lambda sf$. The spin channels/interconnects may require buffer/repeaters based on the channel length and the fanout associated with each output nanomagnet. In our analysis, we ignored any buffers required for charge based interconnects and only analyzed the buffer insertion for spin-channels to understand the design issues related to spin logic.

We estimate the spin channel lengths (and hence, the buffers) by placement of different ASL logic blocks. At first a layout of each nanomagnet was obtained and logic gates and logic functions were placed based on the layout of a nanomagnet. Figure 16 and 17 show a layout of a nanomagnet and a NAND gate respectively [37].

The placement is performed utilizing a hierarchical approach. Initially, a square layout consisting of a 3x3 grid is considered. Then, we partition the circuit into nine different partitions utilizing the *hMetis* partitioning tool [38,39]. Subsequently, the inter-partition connectivities are analyzed to determine a coarse floorplan of the circuit. Each partition is placed in one of the nine sections in the grid based on the connectivities between the partitions. The number of levels of hierarchy is determined based on the size of the circuit. The same steps are repeated hierarchically: each partition is again partitioned and inter-partition connectivites are analyzed to obtain a placement of sub-blocks. Once the placement is performed, the Half Perimeter Wire Length (HPWL) [40] is used to obtain the length of each wire and if the wire exceeds the spin diffusion length, an interconnect buffer is inserted. Thus, an approximate number of required interconnect buffers is obtained. Finally, the netlist is updated using the list of interconnect buffers (and corresponding gating transistors). Note that stacking (and pipelining for combinational circuits) is again performed to obtain a post-placement optimization of the circuit.

## IV. RESULTS AND ANALYSIS

We utilized the proposed design methodology to synthesize ASL implementations for three different classes of circuits, *viz.* random combinational and sequential logic circuits, digital signal processing data-paths and a general-purpose processor (Leon SPARC3). In this section, we present and analyze the results and study the feasibility of ASL as an alternative to CMOS.

### A. Experimental Methodology

Table 1 shows the device parameters of ASL nanomagnets utilized in the physics based simulation framework described in Section III A. These device parameters can be achieved with the current state-of-the-art [27,41,42]. As a

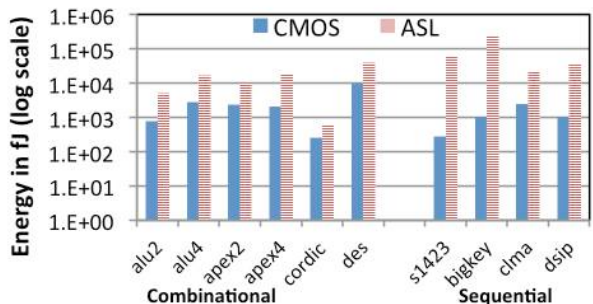
Fig. 18: Energy of ASL and CMOS implementations at 160 MHz.

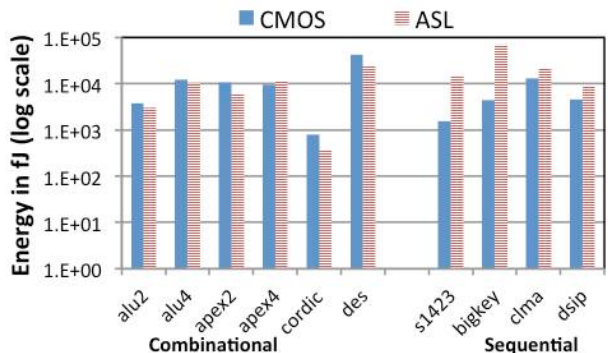
Fig. 19: Energy of ASL and CMOS implementations at 25 MHz.

TABLE I: DEVICE PARAMETERS OF ASL NANOMAGENTS

| Parameter | Value |
| --- | --- |
| Damping factor ($\alpha$) [42] | 0.01 |
| Magnet Volume (V) | 48x16x1 nm$^3$ |
| $K_uV$ | $20K_BT$ |
| $M_S$ [41] | 800 emu/cm$^3$ |
| $H_k$ | 12.272 kOe |
| HighP/LowP | 0.7/0.1 |
| Channel Materials | Cu |
| Spin Diffusion length($\lambda sf$) [27] | 500nm |
| Channel Resistivity | 7 $\Omega$-nm |

representative result, an ASL inverter requires 500 µA of current for the duration of 2.5ns. The delay of ASL gates can be reduced by suitably increasing the supply current. The current is supplied to each nanomagnet using a transistor (we assumed the 16 nm technology node), with its gate terminal connected to a multi-phase gating clock of the desired operating frequency. In our evaluation, we assume the maximum number of nanomagnets that can be stacked in series in the ASL implementation to be 6. This assumption is conservative and does not considerably increase the complexity of power routing, as these nanomagnets are invariably part of at most 2-3 logic gates.

Our ASL technology library consists of 30 logic gates, with a maximum gate fan-in of 7. Each of these gates was designed and characterized using the device parameters presented in Table 1. We utilized Synopsys Design Compiler to obtain an initial mapping of the target design to the ASL library and developed custom tools to automatically perform the various optimizations described in Section III, *viz.* fan-out optimization, intra-cycle power gating, nanomagnet stacking, fine-grained logic pipelining and approximate placement and interconnect estimation to insert the required interconnect buffers/repeaters. We then evaluated design metrics such as area, delay and energy of the resultant ASL implementations. We note that the results include power consumed by all the components in the implementation, including the ASL gates, interconnect buffers, and power gating transistors.

Our benchmark suite comprised of three classes of circuits. The first class was random combinational and sequential logic circuits from the MCNC benchmark suite [43]. The second class of circuits was data-paths from the Digital Signal Processing (DSP) domain *viz.* 16-tap FIR filter and 8-input 1D-DCT. The final benchmark was the open source implementation of Leon SPARC3 general-purpose processor [44]. We compared the designs at two different performance targets – a moderate operating frequency of 160 MHz and a low frequency of 25 MHz. The lower frequency of 25MHz reflects the regime in which ultra-low power ICs used in sensors and implantable devices operate. On the other hand, the higher frequency was limited to 160MHz because ASL faces reliability challenges due to the large currents required to switch nanomagnets at higher speeds.

We would like to clarify the relationship between the operating frequency of the overall circuit (25/160 MHz) and the gating clock used to switch the gating transistors in each gating domain. If the circuit is not pipelined, the width of the clock pulse supplied to each gating domain should be lower than the target clock period of the circuit divided by the number of the gating domains in the implementation. This is because in each "cycle" of operation, all the gating domains in the circuit need to be evaluated sequentially. In the cases when fine-grained logic pipelining is employed, the gating clock pulse width is less than half the target clock period, due to the two-phase pipelining scheme that requires 2 (odd and even) gating domains. Thus, the maximum operating frequency of a circuit depends on the number of gating domains in the implementation. As mentioned above, we do not target operating frequencies beyond 160 MHz, as some of the benchmarks with large numbers of gating domains cannot operate reliably due to excessive current requirements.

For the CMOS baseline, we used the 16nm PTM technology library [45] and the designs were optimized for energy using Synopsys Design Compiler. Synopsys Power Compiler and Synopsys Primetime were used for power, area and timing evaluation. It is worthy noting that the CMOS baseline is well optimized, including gate sizing and voltage scaling in the case of low frequency operation.

### B. Energy and Area Comparison

In this subsection, we compare the area, performance and energy consumed by ASL circuits with CMOS implementations for different benchmark categories. We note that the results presented in this section are at the logic level and do not consider physical design information, *i.e.,* the interconnect buffers and their overheads are not included. The impact of interconnect buffers on energy and area is described in Section IV.C.

*1. Random logic:* Figure 18 compares the energy consumption of ASL to the 16nm CMOS baselines for random combinational and sequential benchmarks operating at 160 MHz. We observe that, despite the proposed optimizations, the energy of ASL is 4-10X higher than CMOS in the case of combinational logic and two to three orders of magnitude higher for sequential logic. The primary reason behind the energy inefficiency of ASL is the large current required for switching the nanomagnets at low delays. It is worth noting that although the overall circuit operating frequency is 160 MHz, individual nanomagnets are switched at a higher frequency based on the number of gating domains in the implementation. Also, despite power-gating (Section III.C), the power consumed by the ASL implementation is still significant. This is largely due to two factors. First, while power-gating nanomagnets reduces the period during which power is consumed, the power consumed during evaluation is still high. Second, ASL gates consume power (during their evaluation period) irrespective of whether the output nanomagnet switches from its current state. This is unlike CMOS, where the energy consumed is a strong function of the switching activity in the circuit. Since the switching activity observed in typical circuits is quite low (~0.1-0.2), a significant fraction of the overall energy consumed by ASL is attributed to power consumed even when gates' outputs do not change.

As observed in Figure 18, ASL is more efficient at realizing combinational circuits since they can be pipelined in a fine-grained manner. For a given throughput, pipelining allows each stage to operate at a lower delay, requiring lower current for evaluation. On the other hand, general sequential circuits are not amenable to fine-grained pipelining due to the inherent cyclic dependencies in their structure, leading to ASL performing significantly worse for them.

Figure 19 compares the energy consumption of ASL and CMOS implementations of combinational and sequential benchmarks operating at 25 MHz. In this case, we observe that the combinational implementations are competitive with CMOS (10% lower energy on average), while the sequential circuits are still worse by an order of magnitude. The improved efficiency of ASL implementations at lower frequency is attributed to the decrease in the magnitude of current that needs to be passed through each nanomagnet. This is complemented by the fact that the energy benefits due to voltage scaling of CMOS implementations subside beyond a certain point, as the delay of CMOS gates begins to grow exponentially with lower supply voltage.

Finally, Figure 20 presents the area of ASL and CMOS implementations. We find the ASL implementations to be, on an average, ~8X better compared to CMOS.

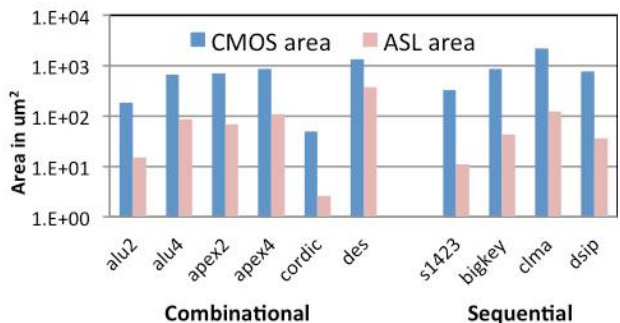
Fig. 20: Area of ASL and CMOS for different benchmarks.

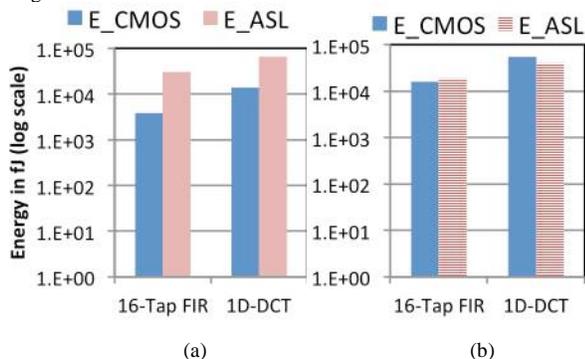
(a)            (b)
Fig. 21: Energy of ASL and CMOS implementations for 16-tap FIR and 1D-DCT data-paths at (a) 160 MHz and (b) 25 MHz.

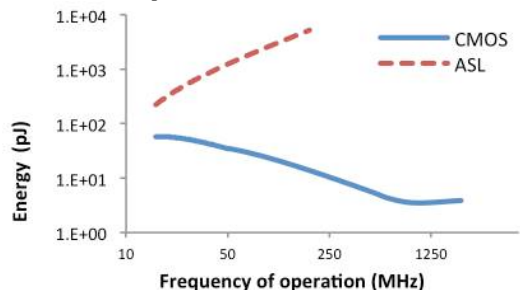
Fig. 22: Energy vs. delay sweep for Leon SPARC3 processor.

*2. DSP Data-paths:* Previous research efforts [28,31] have demonstrated that ASL is efficient for implementing arithmetic circuits such as adders and multipliers. Therefore, we investigated more complex DSP data-paths that are dominated by arithmetic components. Figures 21 (a) and (b) show the normalized energy of CMOS and ASL implementations for the 16-tap FIR and 1D-DCT circuits at both target operating frequencies.

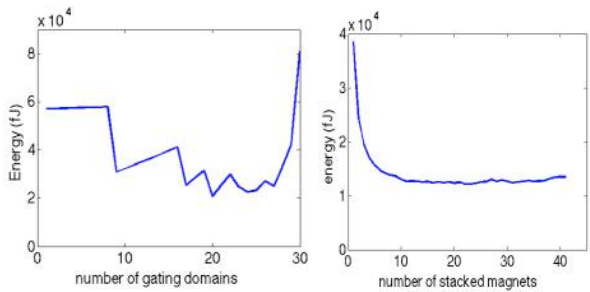
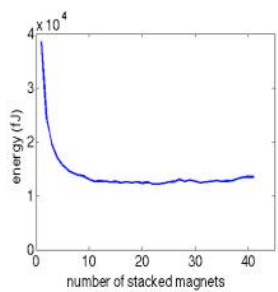
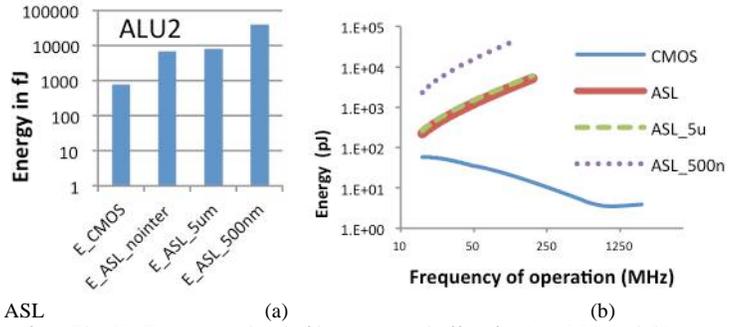

Fig. 23: Energy *vs.* number of gating domains for ALU2 benchmark.

Fig. 24: Energy consumption of ASL circuits with nanomagnet stacking for ALU2 benchmark.

Fig. 25: Energy overhead of interconnect buffers for (a) ALU2 and (b) Leon SPARC3 benchmarks for 500 nm and 5um spin diffusion length.

We observe that, while the ASL implementation is energy inefficient at 160 MHz, it provides a small energy improvement of 10% at 25 MHz for the DCT benchmark. The improvement in energy can be attributed to the ability to pipeline the DSP circuits in a fine-grained manner.

*3. Leon SPARC3 General Purpose Processor:* Finally, we evaluate ASL in the context of a general-purpose processor. In this case, instead of comparing CMOS and ASL implementations at just two performance targets, we perform a complete sweep of the entire design space between 2 GHz to 15 MHz and present the results in Figure 22.

First, in the case of CMOS, as the frequency is decreased, the energy of the implementation also decreases due to the corresponding scaling of the supply voltage. However, beyond a certain point, the exponential growth in delay outweighs the power benefits due to the reduction in voltage, leading to a net increase in energy. In the case of ASL, it is not possible to reliably operate beyond a frequency of 160 MHz due to the large switching current requirement. We see that at 160MHz, the energy consumption is 3 orders of magnitude higher than CMOS. However, as the frequency is decreased, the overall energy of the ASL implementations proportionately decreases as smaller currents are used for ASL evaluation. At the lowest frequency of 15 MHz, the optimized ASL is still ~5X worse than CMOS. The Leon SPARC3 is an in-order pipelined processor and, while ASL can be used to efficiently implement the flip-flops in the circuit, it is not amenable to more fine-grained pipelining as feedback paths between the different pipeline stages cause complex structural dependencies in the logic. The area of the ASL implementation is 3X lower than that of CMOS.

*C. Impact of Different Optimization Steps*

In this subsection, we study the impact of the different optimization steps *viz.* intra-cycle power gating and nanomagnet stacking, on the energy consumption of the ASL implementation using the combinational benchmark circuit ALU2.

Intra-cycle power gating minimizes the static power consumed by the nanomagnets by switching them ON only for the duration when they are required for logic evaluation. However, as outlined in Section III C, it involves energy overheads in terms of gating transistors and multi-phase clocks. Thus, it is critical to identify and partition the circuit into gating domains such that the nanomagnets are switched ON for the least amount of time, while incurring minimal energy overheads. Figure 23 plots the energy consumption for varying number of gating domains used in the implementation for the ALU2 benchmark. We observe that, when the number of gating domains is small, the large static power consumed by the nanomagnets leads to increased energy consumption. When the number of domains is increased, the energy consumption decreases due to reduction in static power. However, beyond a certain point, the energy overheads of the gating transistors become significant, outweighing the benefits of static power reduction. Given a circuit, our methodology identifies the right number of gating domains by analyzing this trade-off.

Next, we analyze the trade-off between stacking ASL nanomagnets and the energy consumption of the circuit. As mentioned in Section III D, when more nanomagnets are connected in series, the energy consumption reduces proportionately due to the orders of magnitude difference in the resistances of the nanomagnets and the gating transistor. However, since all nanomagnets in the stack are evaluated together, the time duration for which the nanomagnets in the stack are powered ON (union of the time duration of all stacked nanomagnets) may increase if the inputs to the nanomagnets arrive at different times. This results in an increase in the energy consumption of the circuit.

Figure 24 plots the energy *vs.* the degree of stacking for ALU2 benchmark. When the number of stacked nanomagnets is low, the energy decrease due to the increase in the degree of stacking is linear. However, when the number of stacked nanomagnets increases beyond a certain point, the energy improvements saturate, as the increase in the evaluation period nullifies the benefits due to stacking. The degree of stacking at which the benefits saturate may differ across circuits. For example, in the case of the Leon SPARC3 processor, a stacking degree of 30 yields 2X improvement over the stacking degree of 15 considered in Figure 22. This is because the processor contains a large number of nanomagnets with similar time periods of evaluation and hence the increase in power ON duration for larger degrees of stacking is negligible. Given a maximum degree of stacking, our synthesis methodology evaluates this trade-off to automatically identify the optimal number of stacked nanomagnets that minimizes the overall energy consumption of the circuit.

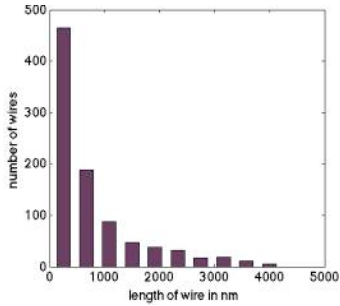
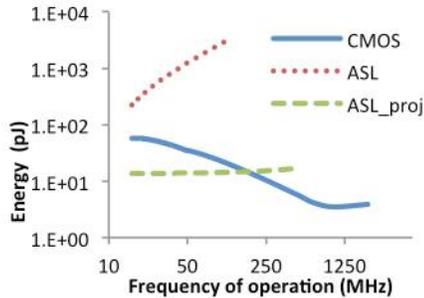

Fig. 26: Histogram of the wire length for ALU2 benchmark.

Fig. 27: Energy of Leon SPARC3 processor with projected parameters for ASL nanomagnets.

TABLE II: CURRENT AND PROJECTED DEVICE PARAMETERS OF ASL

| Parameter | Current | Projected [46-48] |
|---|---|---|
| Damping factor ($\alpha$) | 0.01 | 0.007 |
| Magnet volume (V) | 48x16x1 nm$^3$ | 15x15x1 nm$^3$ |
| $K_u V$ | $20 K_B T$ | $20 K_B T$ |
| $M_S$ | 800 emu/cm$^3$ | 300 emu/cm$^3$ |
| $H_k$ | 12.272 kOe | 12.272 kOe |
| HighP/LowP | 0.7/0.1 | 1.0/0.1 |
| Spin diffusion length ($\lambda$) | 500nm | 10µm |
| Channel resistivity | 7 Ω-nm | 7 Ω-nm |

### D. Impact of Interconnects on ASL Energy

One of the key design constraints to ensure correct functionality of ASL circuit implementations is to account for the limited $\lambda sf$. As mentioned in Section II, if the interconnect length between ASL gates is larger than $\lambda sf$, then appropriate number of buffers need to be inserted in order to reduce their length. In this subsection, we present the energy overheads of interconnect buffers obtained using the approximate place-and-route strategy described in Section III F.

Figure 25 shows the energy of ASL implementations including the overheads of interconnect buffers for two benchmark circuits *viz.* ALU2 and Leon SPARC3. In the case when $\lambda sf$ is 500nm (Cu), we observe that the energy with the interconnect overheads is ~8X compared to the ASL implementation without the overheads. However, when the diffusion length is relaxed to 5um as the case with GaAs at low temperature [46], the interconnect overheads are almost negligible. To gain further insight into the source of this overhead, Figure 26 shows the wire length distribution for the ALU2 benchmark. We find that over 50% of the wires require at least one interconnect buffer. The number of interconnect buffers account for over 70% of the nanomagnets in the circuit.

In addition to the direct energy contribution of the additional buffers, interconnects indirectly impact the energy consumption of ASL in several ways. For example, the long interconnects increase the logic depth of the ASL implementation as they require a chain of buffers to be inserted. Larger logic depth leads to an increased number of gating domains, which implies that for a given target frequency, the individual gating domains need to be operated at a higher frequency. This results in the entire circuit being evaluated using a larger current, and thereby causes significant energy overhead. Also, if the ASL implementation is pipelined, larger number of gating domains leads to further additional buffers to balance all the execution paths. Thus, the $\lambda sf$ of the channel material can have a pronounced effect on the overall energy consumption of ASL circuits.

### E. Projections of Improvements in Device Parameters on ASL Efficiency

From the above evaluations, we identify that the strength of ASL is its non-volatility, which enables it to realize sequential elements with negligible overhead. However, the large switching current required to operate nanomagnets at high frequencies, the static power consumed in the nanomagnets during evaluation and small $\lambda sf$ are key bottlenecks to the energy efficiency of ASL. Clearly, ASL needs to be further optimized (through materials, device and circuit optimizations) to address these bottlenecks. In this section, based on various experimental studies [46,47], we project the impact of improvements in different device parameters of that could facilitate ASL to be competitive with CMOS.

Table 2 shows the current and projected nanomagnet parameters. By technology scaling, the nanomagnet size can be reduced; furthermore, by utilizing lower saturation magnetization ($M_S$) reported in [47], the nanomagnets can be switched more efficiently. This leads to smaller evaluation current for a given target frequency. Also, improving the spin injection efficiency (High-P) from the nanomagnet to the channel material leads to larger spin torque at the output nanomagnet of the ASL gate resulting in better switching characteristics. Finally, $\lambda sf$ can be significantly improved by employing a different channel material such as Graphene [48]. Note that several challenges do exist in integrating/processing these materials/parameters into spin-based logic at this time. However, work has started in earnest to experiment with different channel and magnetic materials and their interfaces to improve the efficiency of ASL.

Based on the above projected parameters, Figure 27 shows the energy consumption of the Leon SPARC3 processor at different performance targets. We find that the ASL is still not suited for high frequency operation. However, as the frequency is reduced beyond a certain point, ~200 MHz in this case, the projected ASL becomes competitive to CMOS and the improvement in energy grows at lower frequencies. At a frequency of 25 MHz, the projected ASL is ~8X better compared to CMOS. Also, a decrease in the target frequency does not improve the energy consumption as rapidly as ASL with current values of parameters. This is due to the limitation on the $V_{ds}$ of the gating transistors, which prevents scaling the voltage beyond a certain point.

Another avenue to improve the efficiency of ASL is to exploit its lower area at higher levels of abstraction. For example, the reduction in area can be used to increase the degree of parallelism (e.g. number of cores/processing elements) in the implementation, benefiting both performance and energy.

## V. CONCLUSION

All Spin Logic (ASL) is a recently proposed logic style that utilizes Spin Transfer Torque (STT) devices to realize

Boolean logic circuits. While the devices possess several favorable characteristics such as non-volatility, high density and the ability to operate at low voltages, their suitability to realize arbitrary logic functions remains hitherto unexplored. Towards this end, we proposed an automatic synthesis methodology to design logic circuits using ASL. The design methodology implements three key optimizations *viz.* intra-cycle power gating, nanomagnet stacking and fine-grained logic pipelining, all of which exploit the unique attributes of ASL to improve its energy efficiency. We utilized the proposed methodology to explore the suitability of ASL for a wide range of benchmarks, including random combinational and sequential logic, DSP data-paths and a general purpose Leon SPARC3 processor and compared ASL to CMOS at the 16nm technology node. We identify that the delay or the switching speed of ASL nanomagnets together with the short spin diffusion length of the non-magnetic channels are key bottlenecks to their efficiency. While logic optimizations such as fine-grained pipelining are targeted to alleviate this bottleneck, they may not be possible in all cases due to the presence of sequential dependencies in the logic. In such cases, the current required to match the performance of CMOS can be quite high, and this significantly impacts the energy-efficiency of ASL. Other optimizations such as nanomagnet stacking also significantly improbe the efficiency of ASL. However, the degree of stacking or the number of nanomagnets that can be stacked together is constrained by the physical limitations of routing. In summary, we conclude that ASL might show some promise for specific classes of low-power/frequency applications such as biomedical applications, internet of things *etc*. Significant improvements in the device switching times and longer spin diffusion lengths of spin channels are critical for ASL to become competitive to CMOS.

**Acknowledgement**: This research was funded in part by the Center for Spintronics: Materials, Interfaces, and Architecture (C-SPIN), a StartNet center funded by DARPA and SRC and by NSSEFF Fellows program.

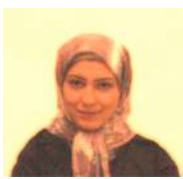
**Zoha Pajouhi (S'04)** is currently working toward the Ph.D. degree in electrical and computer engineering at Purdue University, West Lafayette, IN.
Her research interests include CAD methodology and reliability analysis for beyond CMOS and spin-based devices. She is also interested in VLSI design for signal processing and communication systems.
Zoha received her B.S. and M.S. in electrical engineering from Sharif University of Technology and University of Tehran respectively.

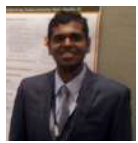
**Swagath Venkhataramani** is a 5th year PhD student in the School of ECE, Purdue University, where he holds the Intel PhD fellowship. Previously, he graduated with a Bachelors degree in Electrical and Electronics Engineering from College of Engineering Guindy, Anna University, Chennai, India in 2010. His research interests include Approximate computing, Energy efficient system-on-chip design and Electronics design automation.

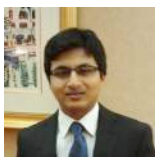
**Karthik Yogendra** is currently pursuing his PhD at the School of Electrical and Computer Engineering, Purdue University, under the guidance of Prof. Kaushik Roy. His broad research interests include Boolean and non-Boolean computation using spin based devices. His focus is mainly on simulation and modeling of Coupled Spin Torque Nano Oscillators for various low power applications. He completed Bachelor of Engineering from Sri Jayachamarajendra College of Engineering (SJCE), Mysore, India in 2007 and Master of Technology (M.Tech) from Indian Institute of Technology (IIT) Bombay in 2011. He worked at IBM T.J Watson Research Center, Yorktown Heights, NY in summer 2013 on simulation and analysis of variability of nano-CMOS devices.

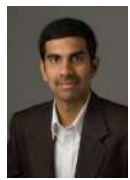
**Anand Raghunathan** is a Professor and Chair of VLSI in the School of Electrical and Computer Engineering at Purdue University, where he leads the Integrated Systems Laboratory. His research explores domain-specific architecture, system-on-chip design, embedded systems, and heterogeneous parallel computing. Previously, he was a Senior Research Staff Member at NEC Laboratories America and held the Gopalakrishnan Visiting Chair in the Department of Computer Science and Engineering at the Indian Institute of Technology, Madras. Prof. Raghunathan has co-authored a book (``High-level Power Analysis and Optimization"), eight book chapters, 21 U.S patents, and over 200 refereed journal and conference papers. His publications have been recognized with eight best paper awards and four best paper nominations. He received the Patent of the Year Award (recognizing the invention with the highest impact), and two Technology Commercialization Awards from NEC. He was chosen by MIT's Technology Review among the TR35 (top 35 innovators under 35 years, across various disciplines of science and technology) in 2006, for his work on "making mobile secure". Prof. Raghunathan has served on the technical program and organizing committees of several leading conferences He was a recipient of the IEEE Meritorious Service Award (2001) and Outstanding Service Award (2004). He is a Fellow of the IEEE, and Golden Core Member of the IEEE Computer Society. Prof. Raghunathan received the B. Tech. degree in Electrical and Electronics Engineering from the Indian Institute of Technology, Madras, and the M.A. and Ph.D. degrees in Electrical Engineering from Princeton University.

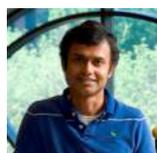
**Kaushik Roy** received B.Tech. degree in electronics and electrical communications engineering from the Indian Institute of Technology, Kharagpur, India, and Ph.D. degree from the electrical and computer engineering department of the University of Illinois at Urbana-Champaign in 1990. He was with the Semiconductor Process and Design Center of Texas Instruments, Dallas, where he worked on FPGA architecture development and low-power circuit design. He joined the electrical and computer engineering faculty at Purdue University, West Lafayette, IN, in 1993, where he is currently Edward G. Tiedemann Jr. Distinguished Professor. Dr. Roy has published more than 600 papers in refereed journals and conferences, holds 15 patents, graduated 65 PhD students, and is co-author of two books on Low Power CMOS VLSI Design (John Wiley & McGraw Hill).
Dr. Roy received the National Science Foundation Career Development Award in 1995, IBM faculty partnership award, ATT/Lucent Foundation award, 2005 SRC Technical Excellence Award, SRC Inventors Award, Purdue College of Engineering Research Excellence Award, Humboldt Research Award in 2010, 2010 IEEE Circuits and Systems Society Technical Achievement Award, Distinguished Alumnus Award from Indian Institute of Technology (IIT), Kharagpur, Fulbright-Nehru Distinguished Chair, DoD National Security Science and Engineering Faculty Fellow (2014-2019), and best paper awards at 1997 International Test Conference, IEEE 2000 International Symposium on Quality of IC Design, 2003 IEEE Latin American Test Workshop, 2003 IEEE Nano, 2004 IEEE International Conference on Computer Design, 2006 IEEE/ACM International Symposium on Low Power Electronics & Design, and 2005 IEEE Circuits and system society Outstanding Young Author Award (Chris Kim), 2006 IEEE Transactions on VLSI Systems best paper award, 2012 ACM/IEEE International Symposium on Low Power Electronics and Design best paper award, 2013 IEEE Transactions on VLSI Best paper award. Dr. Roy was a Purdue University Faculty Scholar (1998-2003). He was a Research Visionary Board Member of Motorola Labs (2002) and held the M.K. Gandhi Distinguished Visiting faculty at Indian Institute of Technology (Bombay). He has been in the editorial board of IEEE Design and Test, IEEE Transactions on Circuits and Systems, IEEE Transactions on VLSI Systems, and IEEE Transactions on Electron Devices.